\documentclass[12pt]{article}

\usepackage{amssymb}
\usepackage{epsfig,color}
\usepackage{pstricks,graphicx,epsfig,color,amssymb,amsmath,amscd}
\usepackage{cite}
\usepackage[]{graphicx}
\usepackage{makeidx}

\newcommand{\be}{\begin{eqnarray}}
\newcommand{\ee}{\end{eqnarray}}

\usepackage[]{caption}
\renewcommand\rho{\varrho}
\begin{document}

\begin{titlepage}
\title{ On graviton propagation in curved space-time background}
\author{E.V. Arbuzova$^{a,b}$, A.D. Dolgov$^{b,c}$, L. A. Panasenko$^{b}$}

\maketitle
\begin{center}
$^a${Department of Higher Mathematics, Dubna State University, \\Universitetskaya ul. 19, Dubna 141983, Russia}\\
$^b${Department of Physics, Novosibirsk State University, \\Pirogova 2, Novosibirsk 630090, Russia}\\
$^c${Bogolyubov Laboratory of Theoretical Physics, Joint Institute for Nuclear Research,
Joliot-Curie st. 6, Dubna, Moscow region, 141980 Russia}
%$^c${ITEP, Bol. Cheremushkinskaya 25, Moscow 117218, Russia}

%\emailAdd{arbuzova@uni-dubna.ru}
%\emailAdd{dolgov@fe.infn.it}
%\emailAdd{akshalvat01@gmail.com}

\end{center}

%\title{Title}
%\author{}

%\begin{document}

%\maketitle

\begin{abstract}

{Equation describing propagation of gravitational waves (GW) over arbitrary curved space-time background is analyzed. New terms, which are absent in the conventional homogeneous and isotropic Friedmann cosmology, are found. Some examples of realistic metric, where these new terms manifest themselves, are presented. Possible implications to very low frequency GW are briefly discussed.}
%\end{abstract}
%\thispagestyle{empty}
%\end{titlepage}

\end{abstract}
\thispagestyle{empty}
\end{titlepage}

\section{Introduction \label{s-intro}}

 Gravitational wave (GW) propagation over Minkowski and curved backgrounds was considered in vast details 
 in the literature, see e.g. textbooks \cite{LL-2,mtw,Mukha,Magg,SW,GR-2}. 
 However, in cosmological situation this consideration is confined to the
 conformally flat Friedmann-Le'Maitre-Robertson-Walker (FLRW)  spacetime. 
In the present work we lift this limitation and derive equation of motion of gravitational waves in an arbitrary spacetime metric. 
We show that in the equation for GW propagation over arbitrary space-time
there appear additional terms which are absent in the FLRW case.

Formally in the classical book~\cite{LL-2} the expansion of the proper curvature tensors up to the first order over small tensor
perturbations  $h_{\mu\nu}$ is presented in an arbitrary background metric, see equation 108.4.  However, the
equatoin is immediately
reduced to the empty space with vanishing Ricci tensor, $R_{\mu\nu} = 0$, in order to obtain the canonical wave
equation $ D^2 h_{\mu\nu} = 0$, equation 108.7 of this reference. Nevertheless in ref.~\cite{LL-2} it is explicitly stated
that the non-zero Ricci tensor would modify  eq. 108.7 and may be of interest.

The corresponding section 1.5 of ref.~\cite{Magg} is the exact replica of the book~\cite{LL-2}, see equation 1.172. 
This equation is also applied to the empty space with vanishing energy-momentum tensor $T_{\mu\nu}$.
Quoting ref.~\cite{Magg}: “Outside the matter sources $T_{\mu\nu} =0$ … tells us the $R_{\mu\nu} = 0$”.  
This immediately excludes the equation 1.172 from application to cosmology, except for the trivial case of 
asymptotically high frequency. All interesting effects related to the FLRW metric and moreover to
deviations from the FLRW space-time in this case are washed out. 

It is claimed in ref.~\cite{Magg}
that it is impossible to fix the background metric without introduction of the so called 
“Low” and “High” terms corresponding to slow and fast varying quantities. 
This  condition contradicts the main stream in the literature on GW propagation over the FLRW space-time. 
The background metric in the world accepted papers is simply taken  in the FLRW form and it does not
lead to any problem.

Our paper is a minor (though technically more complicated) modification of the description of the GW propagation
in space-times which differ from the FLRW one with examples of realistic background metric defined analogously 
to the definition of the FLRW background. We have found that there exist
some new terms  in the equations describing the GW
propagation and mixing of the tensor modes with the scalar ones which is absent in the FLRW case.

 It is important to note
 that for the spacetime, which differs from the FLRW one, it is impossible to impose the standard gauge conditions 
valid for Minkowski and FLRW metrics. Because of this complication one can not separate propagation 
of purely tensor mode from 
scalar and/or vector ones. 
The fact that there can exist mixing between scalar and tensor modes in the presence of matter 
and strong anisotropy (like in the Bianchi type I space-time) or inhomogeneity is well known. 
We thank the referee for this comment. However, in this work we made more general derivation not specifying any 
concrete form of metric.

As shown in {textbook} \cite{mtw},  graviton propagation in empty 
but curved space-time is governed by the equation
\be 
D_{\alpha}D^{\alpha}h_{\mu \nu} - 2R_{\alpha\mu\nu\beta}h^{\alpha \beta}=0, 
\label{EoM-Grishch}
\ee
where 
%$R_{\alpha\mu\nu\beta}$ is the Riemann tensor and
$h_{\mu \nu}$ is the tensor perturbation of the total metric: 
\be
\bar g_{\mu \nu} =  g_{\mu \nu} + h_{\mu \nu},
\label{metric-tot}
\ee
 $g_{\mu \nu}$ is the background metric, covariant derivative $D_{\alpha}$ is defined with respect to the background metric,  
 and $R_{\alpha\mu\nu\beta}$ is the Riemann tensor in the background space-time, which is supposed to be nonzero,
while the  Ricci tensor vanishes, $R_{\mu\nu}=0$.
 
 It was argued in  paper  \cite{Grishchuk} that the l.h.s. of Eq.~(\ref{EoM-Grishch}) was not
 changed in the case of the graviton propagation 
 in the homogeneous and isotropic universe described by the canonical FLRW-metric and 
 with space-time filled by ideal liquid. Both the background energy-momentum, $T_{\mu\nu}$, and the first order 
 correction to it, $T_{\mu \nu}^{(1)}$, as well as the Ricci tensor, $R_{\mu\nu}$,
 are supposed to be non-zero. We expand the total energy-momentum tensor
$\overline T_{\mu \nu}$ as: 
 \be
 \overline T_{\mu \nu} =  T_{\mu \nu} + T_{\mu \nu}^{(1)},
 \label{T-mu-nu-tot}
 \ee
 where  
 $T_{\mu \nu} $ is the energy-momentum tensor of the background matter, 
 and $T_{\mu \nu}^{(1)}$ is that  induced by perturbations.

 The energy density of gravitational waves is 
 of the second order in $h_{\mu\nu}$ and
 {is neglected, as are all other second order contributions. It is usually} assumed that $T_{\mu}^{ \nu (1)}$ for the {\it mixed} components is zero.  {This condition depends upon the type  of matter perturbations and, 
as is known, it can be  fulfilled for the ideal fluid.}
Under this condition Eq.~(\ref{EoM-Grishch}) rewritten in terms of mixed components
preserves the same form both in empty and filled space. 
 
 {There is a subtle point however. If one naively lifts or lowers indices in $T_{\mu}^{ \nu (1)} $ with the background metric,
 one has to conclude that $T_{\mu \nu}^{ (1)} = 0$. If so, the equations for GW propagation in terms of mixed and lower components
 would not be equivalent, for arbitrary and not only for FLRW metric. Indeed,}
 we  show in what follows that if  $T_{\mu}^{ \nu (1)} =0$, then
 $T_{\mu \nu}^{ (1)} \neq 0$ and an account of this fact leads to equivalent equations
 in terms of mixed and lower components {in arbitrary metric, as shown in this work.}
 
 On the other hand, if one assumes that  $T_{\mu \nu}^{ (1)}= 0$,
 the resulting equation in the filled space would significantly differ, {even in FLRW space-time,}
 from the accepted in the literature
 canonical one written in terms of the mixed components $h^\mu_\nu$, 
 {to say nothing about an arbitrary space-time considered in this paper.}

{In the conventional approach to the description of the gravitational wave propagation 
over FLRW background the following 
transversality and traceless conditions on $h^\mu_\nu$ are imposed:}
\be
D_\mu h^\mu_\nu = 0,\,\,\, {\rm and}\,\,\,\, h_\mu^\mu =0.  
\label{D-h-mu-nu}
\ee
It is verified in numerous works that Eq.~(\ref{EoM-Grishch}) allows to impose these conditions in
the curved Einstein (empty) space with $R_{\mu\nu} = 0$. It is also  well known that the same is true  
also for the filled FLRW space-time. However, as it is found 
in what follows, these conditions are violated in the space-time with 
an arbitrary metric different from the FLRW one. {Hence the longitudinal or scalar modes
of GW may propagate and  in the general case tensor and scalar modes are mixed.   
This situation is similar }
to the  propagation of the longitudinal mode of the electromagnetic wave  in plasma
and the appearance of non-zero effective photon mass equal to the plasma frequency.  

{This statement is criticised by one of the referees of this work, who quoted Ref.~\cite{deser} as that where
it is proven that no
extra degrees of freedom may appear, namely:
\begin{quote}
%I cannot recommend publication: I don’t see what the point is. Instead I do see some obvious errors. For example, 
"the authors completely misrepresent  Ref.~\cite{deser}
as a main point; that paper is so simple as to be obviously correct. Nowhere is any new DoF suggested 
for any case there–quite the opposite." 
\end{quote}
In Ref.~\cite{deser} a very special case of Ricci-flat metric (i.e. spaces with $R_{\mu\nu} = 0$) is considered. So there is not much sense to conclude that our paper is wrong because it disagrees with Ref.~\cite{deser}.
On the opposite, in the case of the Ricci flat metric with  $R_{\mu\nu} = 0$
our results completely agree with Ref.~\cite{deser}. Moreover we also fully agree with
the numerous existing literature treating GW propagation in FLRW space-time with $R_{\mu\nu} \neq 0$, to
which the results of Ref.~\cite{deser} are not applicable. }

Moreover,  the basic Eq. (2) of Ref.~\cite{deser}
contradicts the accepted canonical equation for GW propagation in FLRW metric if we formally take non-zero Ricci tensor. 
It does not mean of course that all world literature contains ”some obvious errors” quoting 
this referee.
We repeat that we derived equation for GW propagation in arbitrary space-time. In the limit of FLRW metric (note that 
in this metric  $R_{\mu\nu}\neq 0$ ) it coincides with the numerous published works and in the Ricci-flat case it 
agrees with eq. (2) of Ref.~\cite{deser}.
So the conclusion of the referee that our paper is incorrect is erroneous.

The noticed above {an apparent}
inconsistency between equations describing GW propagation in terms of $h_{\mu\nu}$ 
and $h_\nu^\mu$ is related to the fact that  $T^{(1)}_{\mu\nu} \neq g_{\mu\alpha} T^{(1) \alpha}_\nu$. Indeed
\be
\bar T _{\mu \nu} = \bar g _{\alpha\nu} \bar T_{\mu}^{\alpha} = 
\left( g_{\alpha \nu} + h_{\alpha\nu} \right) \left( T^{\alpha}_{\mu} + T^{(1)\alpha}_{\nu} \right), 
\label{T-mu-nu-1}
\ee
so $T_{\mu\nu}^{(1)} = h_{\alpha\nu} T^\alpha_\mu + g_{\mu\alpha} T^{(1) \alpha}_\nu$.
The addition of this term into the r.h.s of Eq.~(\ref{EoM-Grishch}) allows to impose the transversality condition
on this equation, so the extra  propagating modes are not excited over FLRW background
in contrast to the concern expressed
in Ref.~\cite{deser}. For more detail see the discussion below Eq.~(\ref{bar-T-mu-nu}).

In textbook \cite{GR-2} the detailed derivation of the equation 
{describing GW propagation over FLRW background}, 
which is fully equivalent to
 Eq.~(\ref{EoM-Grishch}), is presented for the background interval 
 rewritten in terms of conformal time $\tau$:
 \be
 ds^2 = a^2(\tau)\left(d\tau^2 - \delta_{ij}dx^i dx^j\right)
 \label{ds-conf} 
 \ee
under assumption that $a(\tau)$ is a scalar with respect to coordinate transformation. Here $\delta_{ij}$ is the Kronecker 
symbol.  The transition to conformal time is convenient to make for conformally flat
FLRW metric, but in the case of an arbitrary space-time this transition is not particularly useful because
a metric, which is not conformally flat, cannot be transformed to that proportional to the Minkowski metric.

In this paper we generalise Eq.~(\ref{EoM-Grishch}) for graviton propagation over arbitrary space-time 
background. We show that in the l.h.s. of this equation 
there appear {additional terms, which vanish} in FLRW metric and which can 
dominate in the low frequency limit. 

{Remind that the study of metric perturbations was pioneered in ref.~~\cite{lifshitz-pert}, 
see also~\cite{lif-khal}, where } {it is shown that}
$h_{\mu\nu}$ could be separated into three types
according to their properties with respect to 3-dimensional space 
transformation (rotation): scalar ($h_{t}^t$ and $h^{i}_i)$, 
vector ($h_t^i $, $\partial_i h^i_t  =0$), and tensor ($h_i^j$). The coordinate freedom allows to impose the
following conditions on $h_i^j$:
\be
\partial_i h^i_j = 0, \,\,\, h_i^i =0 ,
\label{gauge-cond}
\ee
in locally Minkowsky frame. These conditions can be imposed also globally in conformally flat
FLRW metric.
So the tensor mode $h_i^j$ has 
only two independent components describing massless quanta propagation with two, as it should be,
transverse helicity states, i.e. gravitational waves. However, in general space-time this  might be incorrect 
and additional propagating degrees of freedom could be excited. { The problem of gauge fixing in arbitrary space-time
is considered in Sec.~\ref{S-gauge}.} 

Usually these perturbations are  considered either over vacuum solutions of the Einstein equations or
over background FLRW space-time with the metric:
\be
ds^2_F = dt^2 - a^2(t)\delta_{ij}dx^i dx^j ,
\label{ds-FLRW}
\ee
In what follows we assume for simplicity  that the FLRW metric is spatially flat.
Transition from this metric to conformal one (\ref{ds-conf}) is given by the linear 
{transformation between time components of the coordinates,
$dt = a(\tau ) d\tau$.

{Gravitational wave production {and propagation
in FLRW-}cosmology was studied in Refs.~\cite{Grishchuk,zakharov-79,StarGW,VerRubSazh}.   
According to the Parker theorem, see Ref.~\cite{parker1, parker2}, massless particles production by 
conformally flat space-time, 
metric, {such as, in particular,} FLRW metric is forbidden  
if the corresponding field equations are conformally invariant. This is true for massless fermions, conformally 
coupled scalars,
and massless vector fields, up to possible breaking of conformal invariance by the trace 
anomaly~\cite{ad-conf-anom}.
It was discovered by L. Grishchuk that gravitons can be produced in conformally flat space-time since their 
equation of motion is not conformally invariant \cite{Grishchuk}. 
Gravitational waves could be efficiently produced during cosmological inflation \cite{StarGW,VerRubSazh}. 
They may be observed by the cosmic interferometer LISA  which will presumably reveal
information about the  mechanism of the primordial inflation. }

The paper is organised as follows. 
Sec.~\ref{S-gauge}  is devoted to the choice of gauge for 
{four-dimensional tensor modes in arbitrary space-time and comparison that with the popular gauges in 
FLRW metric.}
In Sec.~\ref{s-bas-eq} we derive equations describing propagation of
gravitational waves   in an arbitrary spacetime making expansion of the exact Einstein equations up to the first 
order metric perturbations $h_{\mu\nu} $.  In Sec.~\ref{s-t-mu-nu-1} the problems with the first order corrections 
to the energy-momentum tensor are discussed. Further, in Sec.~\ref{s-mixed} 
the equation for the propagation of the mixed components, $h^\mu_\nu$,   is derived.
Section \ref{s-trans} is devoted to the proof that the transversality condition 
{ $D_\mu (h^\mu_\nu - \delta^\mu_\nu h^\alpha_\alpha/2)= 0$}
 is compatible with the equation of GW propagation and so it can be imposed in arbitrary metric.
In Sec.~\ref{s-non-FLRW} a few examples of realistic metrics, different from
FLRW one,  are presented. Lastly, in Sec.~\ref{s-concl} possible implications of our results to low frequency GWs
are briefly discussed.

{ \section{Choice of gauge} \label{S-gauge}

{ To avoid confusion expressed by one of the referees we stress again that  below we consider an 
arbitrary space-time metric but not only the  FLRW one. In the latter case the choice of gauge is heavily 
based on particular properties of FLRW metric.
}

We discuss the choice of gauge conditions which can be imposed on $h_{\mu\nu}$ 
{following the lines presented in} the textbook \cite{LL-2}. 
Let us make the coordinate transformation  $\tilde x^\mu = x^\mu + \xi^{(1)\mu}$, where $\xi^{(1)\mu}$ is  a small vector. 
Under such transformation the first order perturbations, $h_{\mu \nu}$, of the metric \eqref{metric-tot} changes as:
\be 
\tilde h_{\mu \nu} = h_{\mu \nu} - D_\mu \xi^{(1)}_\nu - D_\nu \xi^{(1)}_\mu.
\label{tilde-hmn}
\ee  
Using freedom in the  choice of four functions $\xi^{(1)\mu}$ we can impose the following four conditions:
\be
D_\mu \psi^{\mu}_\nu = 0, \ \ \ 
\label{Dpsi} 
\ee
where $\psi^\mu_\nu = h^\mu_\nu -  \delta^\mu_\nu\, h/2$ and  $h = h^\alpha_\alpha$. In the flat spacetime case 
condition \eqref{Dpsi} leads to 
the wave equation in the classical form, $D^2 h^\mu_\nu =0$. The same equation is true the high frequency (eikonal)
limit.

There remains freedom in the coordinate transformation  
$\tilde h^{\mu}_{\nu}=h^{\mu}_{ \nu} - D^\mu \xi^{(2)}_\nu - D_\nu \xi^{(2)\mu}$ 
with the new parameters $\xi^{(2)\mu}$ which do
not violate condition \eqref{Dpsi}. Accordingly parameters $\xi_\mu^{(2)}$ should satisfy the equation:
\be
D^2 \xi^{(2)}_\mu + R_\mu^\nu \xi^{(2)}_\nu = 0.
\label{D2-xi2} 
\ee
Thus we are allowed to use four more functions $\xi_\mu^{(2)} $ to fix the gauge.

This freedom was used in the textbook~\cite{LL-2} {to impose the restrictions
$h_{ti}=0$ and $h=0$, where $i$ is the space index. 
We, however, apply this freedom to demand $h_{t\alpha}=0$
for any $\alpha $}. In this case the condition $h=0$ may be invalid.

There is still some freedom to make the coordinate transformation with parameter $\xi^{(3)\mu}$, which, 
{ in addition to (\ref{D2-xi2}),} satisfies the condition:
\be 
D_\mu \xi^{(3)\mu} = 0. 
\label{D-xi3}
\ee  
Evidently, the transformation with the functions $\xi^{(3)\mu}$ does not change the value of $h$.

Detailed discussion  of different types of perturbations (scalar, vector, and tensor) can be found e.g. in the book by Mukhanov~\cite{Mukha}.
However, all that {was done there only for the} FLRW spacetime.

\vspace{5mm}
\section{Basic Equations \label{s-bas-eq}}

We start from the exact Einstein equations:
\be
\bar R_{\mu \nu} - \frac{1}{2} \bar g_{\mu \nu} \bar R = 
%\frac{ 8 \pi}{m_{Pl}^2}\, \overline T_{\mu \nu} \equiv 
{\overline {T}_{\mu \nu}}.
\label{EinEq-tot}
\ee 
Here and in the rest of the paper  $\bar T_{\mu\nu}$ and $T_{\mu\nu}$ are related to the physical energy-momentum 
tensor through the constant factor as:
\be
T_{\mu\nu} =\frac{ 8 \pi}{m_{Pl}^2} T^{(phys)}_{\mu \nu} .
\label {T-phys}
\ee

Overline  means that the corresponding exact (total)
quantities are calculated in terms of the total metric $\bar g_{\mu \nu}$, 
see Eqs.~(\ref{metric-tot}) and (\ref{T-mu-nu-tot}).
We will consider first-order tensor perturbations $h_{\mu \nu}$ over background metric $g_{\mu \nu}$ 
and expand the total Ricci tensor and energy-momentum tensor as
\be
\bar R_{\mu \nu} = R_{\mu \nu} + R_{\mu \nu}^{(1)}, \ \ \ \ 
{\overline {T}_{\mu \nu}} =  {T}_{\mu \nu} +  {T}_{\mu \nu}^{(1)},  
\label{Rc-EMT-tot}
\ee
 assuming that all background quantities are taken in the background metric $g_{\mu \nu}$.
 
Our goal is to derive the first-order perturbation equation governing evolution of $h_{\mu \nu}$ without any assumptions about  the form of the background metric. 

From the condition
\be 
\bar g_{\mu \alpha} \bar g^{\nu \alpha} = \delta_{\mu}^{\nu} 
\label{bar-g-inv}
\ee 
it follows that:
\be
\bar g^{\mu \nu} = g^{\mu \nu} - h^{\mu \nu}.
\ee

For the Ricci scalar we have:
\be 
\bar R = \bar g^{\alpha \beta} \bar R_{\alpha \beta} = 
\left(g^{\alpha \beta} - h^{\alpha \beta}\right)\left(R_{\alpha \beta} + R_{\alpha \beta}^{(1)}\right) = 
g^{\alpha \beta}R_{\alpha \beta} - h^{\alpha \beta}R_{\alpha \beta} + g^{\alpha \beta} R_{\alpha \beta}^{(1)}\,.
\label{bar-R-1}
\ee
{ Let us note, that the first order correction to the curvature scalar is not simply obtained from the first order correction to the
Ricci tensor through the contraction of indices by the background metric, but contains an extra term:
\be 
R^{(1)} = g^{\alpha \beta} R_{\alpha \beta}^{(1)} - h^{\alpha \beta}R_{\alpha \beta}.
\label{R1}
\ee} 
Following the book by Landau and Lifshitz \cite{LL-2} we express  the perturbation of the Ricci tensor, 
$R_{\mu \nu}^{(1)}$, via 
metric perturbations, $h_{\mu \nu}$, as
\be
R_{\mu \nu}^{(1)} = \frac{1}{2}\left(D_{\alpha}D_{\nu} h^{\alpha}_{\mu} +
 D_{\alpha}D_{\mu} h^{\alpha}_{\nu} - D_{\alpha}D^{\alpha}h_{\mu \nu} - D_{\mu}D_{\nu}h \right), 
 \label{Ricci-1}
\ee
where covariant derivatives $D_\beta$ are taken with respect to the background metric  $g_{\mu\nu}$ and
$D^{\alpha} = g^{\alpha \beta } D_{\beta}$.
Here and below we use the background metric, $g^{\mu \nu}$ and $g_{\mu \nu}$, to move indices up and down. 

{According to Eq.~\eqref{Dpsi}  
$h^\mu_\nu$ satisfies the condition: 
\be 
D_{\mu} h^{\mu}_{\nu} = \frac{1}{2} \partial_\nu h. 
\label{h-cond}
\ee
}
Using the commutation rules of covariant derivatives we arrive to the result:
\be 
R_{\mu \nu}^{(1)} = - \frac{1}{2}D_{\alpha}D^{\alpha} h_{\mu \nu} +
h^{\alpha \beta} R_{\alpha \mu \nu \beta} + 
 \frac{1}{2}\left(h_{\alpha \mu} R^{\alpha}_{\nu} + h_{\alpha \nu} R^{\alpha}_{\mu}\right).
 \label{Rmn-1}
\ee

Substituting Eqs.~(\ref{Rmn-1}) and (\ref{bar-R-1}) into Eq.~(\ref{EinEq-tot}) and keeping only the first-order quantities
we obtain the following equation for tensor perturbations of the metric:
\be \nonumber
D_{\alpha}D^{\alpha} h_{\mu \nu} - 2 h^{\alpha \beta} R_{\alpha \mu \nu \beta} - 
\left(h_{\alpha \mu} R^{\alpha}_{\nu} + h_{\alpha \nu} R^{\alpha}_{\mu}\right) + 
h_{\mu \nu} R - g_{\mu \nu} h^{\alpha \beta}R_{\alpha \beta} 
{ - \frac{1}{2} g_{\mu \nu} D^2 h} \\
= - 2\, { T}_{\mu \nu}^{(1)}. 
\label{main-eq}
\ee   
{Taking trace we arrive to the equation:
\be 
 D^2 h + 4 h^{\alpha \beta} R_{\alpha \beta} - hR =  2 { g}^{\mu \nu} { T}_{\mu \nu}^{(1)}.  
\label{tr-main-eq}
\ee
In general case $h^{\alpha \beta} R_{\alpha \beta} \neq 0$, %hence as follows from Eq.~\eqref{tr-main-eq} 
so one can conclude that  $h \neq 0$.  
On the other hand in FLRW spacetime the condition $h^{\alpha \beta} R_{\alpha \beta} = 0$ is fulfilled and thus 
equations of motion \eqref{main-eq} and \eqref{tr-main-eq} do not contradict the condition $h=0$, if the source ${ T}_{\mu \nu}^{(1)}$ is traceless.
}

Equation \eqref{main-eq} coincides with the conventional equation (\ref{EoM-Grishch}) in empty space, where $R_{\mu \nu} =0$,
but essentially differs from Eq.~(\ref{EoM-Grishch}) in filled space even in the Friedmann space-time.
Indeed, in FLRW space-time equation (\ref{main-eq})  takes the form:
\be
\left(\partial_t^2 - \frac{1}{a^2}\, \partial^2_k \right)\,h_{ij} - H \partial_t h_{ij} - 2 \left( H^2 + 
3\, \frac{\ddot a}{a}  \right) h_{ij} =  -  2\, { T}_{ij}^{(1)} ,
%-2\, \widetilde T_{ij}^{(1)}, 
\label{main-eq-FRW-2}
\ee
while the conventional Eq.~(\ref{EoM-Grishch})  {with an account of the source term, which we denote as ${ T}_{\mu\nu}^{(1c)}$, } becomes
\be
\left(\partial_t^2 - \frac{1}{a^2}\, \partial^2_k \right)\,h_{ij} - H \partial_t h_{ij} - 2\, \frac{\ddot a}{a} h_{ij} 
=  - 2\, { T}_{ij}^{(1c)} .
\label{conv-FRW-2}
\ee
Surprisingly both equations are true and the resolution of the evident inconsistency is hidden in
a difference between ${T}_{\mu \nu}^{(1)}$ and ${ T}_{\mu \nu}^{(1c)}$, which according to
Eq.~(\ref{T-mu-nu-1}) is 
\be
T_{\mu\nu}^{(1c)} = h_{\alpha\nu} T^\alpha_\mu + g_{\mu\alpha} T^{(1) \alpha}_\nu =
h_{\alpha\nu} R^\alpha_\mu + g_{\mu\alpha} T^{(1) \alpha}_\nu.
\label{T-c-to-T}
\ee
Keeping in mind that propagation of GWs in FLRW background is described under assumption that 
$T^{(1) \mu}_\nu =0$, we can check that the condition $h_\mu^\mu = 0$ is fulfilled in this 
background metric, simply because $h_{\mu\nu} R^{\mu\nu} = 0 $.

\section{First order corrections to  energy-momentum tensor \label{s-t-mu-nu-1}}

The only possible resolution of the discrepancy between Eqs. (\ref{main-eq-FRW-2}) and (\ref{conv-FRW-2})
 lays in the difference between $ { T}_{\mu \nu}^{(1)}$  
and $ { T}_{\mu \nu}^{(1c)}$.
The standard derivation of the conventional
Eqs. (\ref{EoM-Grishch}) or (\ref{conv-FRW-2}) is heavily based on the condition  
$ { T}_{\mu}^{(1c) \nu} =0$, which is incompatible with seemingly equivalent condition
 $ { T}_{\mu \nu}^{(1)} =0$.  Indeed let us assume, that the first  order corrections to the 
 energy-momentum tensor with mixed components are zero,  $T_\mu^{(1) \nu  } = 0$, as it is usually taken
 in the case of FLRW space-time filled by the  ideal fluid. Then, if we consider equation for $h_{\mu\nu}$,
 we should put down indices using the full metric tensor $\bar g_{\mu\nu}$ and so:
\be
\bar T_{\mu\nu} = \bar g_{\mu\alpha} \bar T^\alpha_\nu = 
T_{\mu\nu} + h_{\mu\alpha} T^\alpha_\nu { +  g_{\mu\alpha} T^{(1)\alpha}_{\nu} } = 
T_{\mu\nu} + h_{\mu\alpha} T^\alpha_\nu.
\label{bar-T-mu-nu}
\ee
{ The last equality is true, if $T_\mu^{(1)\nu } = 0$.}

Hence
\be
 T^{(1)}_{\mu\nu} = h_{\mu\alpha} T^\alpha_\nu = 
(1/2) \left( h_{\mu\alpha} R^\alpha_\nu +  h_{\nu\alpha} R^\alpha_\mu -  h_{\mu\nu} R \right),
\label{T-1-mu-nu-low}
\ee
where the Einstein equation for the background curvature is used:
\be 
R_{\mu \nu} - g_{\mu \nu} R/2 =   T_{\mu \nu}.
\label{EinEq-0}
\ee
Substituting expression (\ref{T-1-mu-nu-low}) for $ T^{(1)}_{\mu\nu} $ we obtain from Eq.~(\ref{main-eq})
\be
D_{\alpha}D^{\alpha} h_{\mu \nu} - 2 h^{\alpha \beta} R_{\alpha \mu \nu \beta} 
%\left(h_{\alpha \mu} R^{\alpha}_{\nu} - h_{\alpha \nu} R^{\alpha}_{\mu}\right) 
 - g_{\mu \nu} h^{\alpha \beta}R_{\alpha \beta} {- \frac{1}{2} g_{\mu \nu} D^2 h} = 0 .
\label{main-eq-corr}
\ee   
{ Since the last two terms in this equation vanish} in the FLRW background we arrive to the canonical equation (\ref{EoM-Grishch}).

On the other hand, canonical Eq.~(\ref{EoM-Grishch}) or Eq.~(\ref{conv-FRW-2}), which follows from 
Eq.~(\ref{EoM-Grishch}) for FLRW metric,
is obtained  originally from the mixed component equation for $h_\mu^\nu$. In this case the indices are put down with
the background metric $g_{\mu\nu}$ and thus starting from $T^{\nu (1)}_{\mu} = 0$ we arrive to $T^{(1c)}_{\mu \nu} = 0$,
so both ways ultimately lead to the same result.

Let us note that the condition $T^{\nu (1)}_{\mu} = 0$ is not necessarily true and there are some 
realistic cases when it is not fulfilled. 
For example, $T^{\nu (1)}_{\mu} \neq 0$ in the
equation describing graviton-to-photon transition in external magnetic field even over the Minkowsky background,
see e.g.~\cite{AD-DE}. Another known example is presented by anisotropic stresses which could be 
induced e.g. by neutrinos and photons ~\cite{an-str-1,an-str-2}. They all 
are treated perturbatively, so the background remains the FLRW one.

Let us stress that if the background metric deviates from the Friedmann one, 
the last { two terms in Eq.~(\ref{main-eq-corr})
may essentially change the character of solutions}, especially the first of them} since it does not vanish in zero frequency limit.

\section{Equation for the mixed components \label{s-mixed}}

To make the paper self-contained
let us now derive equation for the mixed components, $h^\mu_\nu$. 
To this end we start from the equation:
\be 
\bar R^{\mu}_{\nu} - \delta^{\mu}_{\nu} \bar R /2 = {\overline {T}^{\mu}_{ \nu}}
%\frac{ 8 \pi}{m_{Pl}^2}\, \overline T_{\mu \nu} \equiv {\overline {\widetilde T}_{\mu \nu}},
\label{EinEq-tot-mix}
\ee 
and decompose it up to the first order:
\be
\bar R^{\mu}_{\nu} = R^{\mu}_{\nu} + R^{ \mu (1) }_{\nu}  
%R_{\mu \nu}^{(1)}, \ \ \ \ 
%{\overline {\widetilde T}_{\mu \nu}} =  {\widetilde T}_{\mu \nu} +  {\widetilde T}_{\mu \nu}^{(1)},  
\label{R-munu-mix}
\ee
The first correction $R^{(1)}_{\mu\nu}$ is calculated in the book of Landau and Lifshitz~\cite{LL-2}
and presented  here in  
Eqs.~(\ref{Ricci-1}) and (\ref{Rmn-1}). The indices are raised according to:
\be
 \bar R_{\nu}^{\mu} =  \bar g^{\beta\mu} \bar R_{\nu\beta} =  
\left( g^{\beta\mu} - h^{\beta\mu }\right) \left(R_{\nu\beta} + R^{(1)}_{\nu\beta}\right)
= R_{\nu}^{\mu} - h^{\mu\beta} R_{\nu\beta} + g^{\mu\beta} R^{(1)}_{\nu\beta}.
\label{bar-r-mu-nu}
\ee
Analogously;
\be
\bar R = \delta^\alpha_\beta \bar R^\beta_\alpha = 
R - h^{\mu\beta} R_{\mu\beta} + g^{\mu\beta} R^{(1)}_{\mu\beta} .
\label{bar-R}
\ee
Finally in the first order we obtain:
\be 
g^{\mu\beta} R^{(1)}_{\nu\beta}  - h^{\mu\beta} R_{\nu\beta} + 
\frac{1}{2} \delta^\mu_\nu h^{\alpha\beta} R_{\alpha\beta} - 
\frac{1}{2} \delta^\mu_\nu g^{\alpha\beta} R^{(1)}_{\alpha\beta} = 0 ,
\label{eq-mix-first}
\ee
where according to the discussion in the previous section we took $\widetilde T^{\mu (1)}_{\nu} = 0$.

%The product $g^{\alpha\beta} R^{(1)}_{\alpha\beta}$ vanishes in any space-time, 
{ In FLRW background the product $g^{\alpha\beta} R^{(1)}_{\alpha\beta}$ vanishes
due to the conditions $D_\alpha h^\alpha_\beta = 0 $ and $h^\mu_\mu =0  $,
and 
$h^{\alpha\beta} R_{\alpha\beta} =0$, since 
$R_{tj} =0$, $R_{ij} \sim \delta_{ij}$,  
  $h^{t \mu} = 0$ and $h_i^i =0$. On the other hand, both these products are generally nonzero if the background
  deviates from the FLRW one.  
 According to Eq.~\eqref{Rmn-1} $g^{\alpha\beta} R^{(1)}_{\alpha\beta}= - D^2 h/2$. As for $h^{\alpha\beta} R_{\alpha\beta}$, it is
  generally non vanishing for an arbitrary taken Ricci tensor.

Taking expression~(\ref{Rmn-1}) for $R^{(1)}_{\nu\beta}$
and using the condition that in FLRW background $ R^i_j \sim \delta^i_j $, we conclude that the last term in Eq. (\ref{Rmn-1})
is equal to $h^{\mu\beta} R_{\nu\beta}$ and cancels out with the second term in
 the l.h.s. of Eq.~(\ref{eq-mix-first}).  Finally, we obtain the conventional equation for the mixed components:
} 
 \be
 D_{\alpha}D^{\alpha} h^{\mu}_{\nu} - 2 h_\alpha^\beta R^{\mu\alpha}_{\,\,\,\,\,\,\,\beta\nu} = 0 ,
 \label{fon-conv}
 \ee
where the indices are lifted with the background metric tensor $g^{\mu\nu}$.

This equation in FLRW background turns into:
\be
\left( \partial_t^2 - \frac{\Delta}{a^2} + 3 H \partial_t   \right)\,h_\nu^\mu = 0.
%- 2 T^{(1) \mu}_\nu.
\label{ex-mixed-FRW}
\ee
Making transformation to $h_{\mu\nu}$ according to
\be
 h^\mu_\nu = g^{\mu \alpha} h_{\alpha\nu} = - h_{\alpha\nu} \delta^{\mu \alpha}/ a^2
 \label{mixed-to-lower}
 \ee
we arrive at
\be 
\left( \partial_t^2 - \frac{\Delta}{a^2}  - H \partial_t  -  \frac{2 \ddot a}{a} \right)\,h_{\nu \mu} = 0.
% 2 a^2T^{(1) \beta}_\nu \delta_{\mu \beta}.
\label{h-lower}
\ee
It coincides with  the conventional Eq. (\ref{conv-FRW-2}) but not with Eq. (\ref{main-eq-FRW-2}) if one assumes that
the right hand sides of both equations vanish.
{ However,  we have shown that taking $T^{\mu (1)}_\nu = 0$ in FLRW metric for mixed components leads to 
$T^{(1)}_{\mu\nu}\neq 0$ with lower indices.}

It may be instructive to present equations in terms of conformal time, since they are frequently
analyzed this way.  However, if the metric is not conformally flat, transition to conformal time generally does not
make much sense.

\section{Verification of the transversality conditions for non-zero $h_\mu^\mu$ \label{s-trans}}

It is shown here that the action of the covariant divergence $D_\mu$ 
%as well as taking the trace from 
on both sides of the equation for $h^\mu_\nu$ 
gives self-consistent results. For easier reading we
repeat here some equations already presented above.

We start from the exact equations
\be
\overline{R}^{\mu}_{\nu}-\frac{1}{2} \delta^\mu_\nu \overline{R}=\overline{{T}}_{\nu}^\mu
%\equiv T^\mu_\nu + T^{(1) \mu}_\nu
\ee
and expand $\overline{{T}}_{\nu}^\mu =T^\mu_\nu + T^{(1) \mu}_\nu$ and 
$\overline{{R}}_{\nu}^\mu = R^\mu_\nu + R^{(1) \mu}_\nu$. So we obtain: for the background metric, as expected:
\be
{R}^{\mu}_{\nu}-\frac{1}{2} \delta^\mu_\nu {R}= {{T}}_{\nu}^\mu.
\label{R-mu-nu-bcgr}
\ee

However, we use the first order correction to $R_{\mu\nu}$
as is presented in Ref.~\cite{LL-2}:
\be
R_{\alpha\nu}^{(1)} = (1/2) \left( D_\beta D_\nu h^\beta_\alpha + D_\beta D_\alpha h^\beta_\nu -
D^2 h_{\alpha\nu}  - D_\alpha D_\nu h  \right),
\label{R1-mu-nu-1}
\ee 
so we  rise one lower index using ${g}^{\mu\alpha}$ and find:
{ 
\be
D^2 h^\mu_\nu  + 2 R^{\mu\alpha}_{.\, .\,\,\nu \beta} h^{\beta}_\alpha + R_{\nu}^{\alpha} h^{\mu}_{\alpha} -
R^\mu_\alpha h_\nu^\alpha - \delta^\mu_\nu \left(R^\alpha_{\beta} h_{\alpha}^{\beta}  + \frac{1}{2} D^2 h \right) =
-2 T^{(1)\mu}_\nu  ,
\label{D2-h-mu-nu-h}
\ee
where $D^2 = g^{\alpha\beta} D_\alpha D_\beta$ and
it is taken into account that  $h=h^{\alpha}_{\alpha} \neq 0$ and 
$D_{\mu}h^{\mu}_{\nu}=\partial_\nu h/2 $.
%and the last term in the r.h.s. is taken from the l.h.s. to make it symmetric with
%respect to $\mu$ and $\nu$ interchange.
% for the last term in brackets in eq. (\ref{first_pert_level}) 
}

Acting by $D_\mu$ on the first term and applying commutation rules for the covariant derivatives  we come to:
{
\be
D_{\mu}D^2 h^{\mu}_{\nu} =
D_{\mu}\left(R_{\sigma}^\mu h^{\sigma}_{\nu}\right)-D_{\mu}\left(R^{\sigma\,.\,.\,\mu }_{\,\,\nu\alpha}h^{\alpha}_{\sigma}\right)-
R^{\sigma.\,.\,\alpha}_{\,\,\nu\mu}D_{\alpha}h^{\mu}_{\sigma}{ + \frac{1}{2} D^2 (D_\nu h)} .
% D^{\alpha}\left(R_{\sigma\alpha}h^{\sigma}_{\nu}-
%R^{\sigma}_{.\,\nu \mu\alpha}h^{\mu}_{\sigma}\right)-R^{\sigma}_{.\,\nu\mu\alpha}D^{\alpha}h^{\mu}_{\sigma} 
%+ \frac{1}{2} D^2 (D_\nu h) .
\label{3cov_der}
\ee
}
%dsp
Substituting Eq. (\ref{3cov_der}) into the divergence of the l.h.s. of 
eq. (\ref{D2-h-mu-nu-h}) we come to the following result
\be
{
D_{\mu}\left(R_{\sigma}^\mu h^{\sigma}_{\nu}\right)-D_{\mu}\left(R^{\sigma\,.\,.\,\mu }_{\,\,\nu\alpha}h^{\alpha}_{\sigma}\right)-
R^{\sigma.\,.\,\alpha}_{\,\,\nu\mu}D_{\alpha}h^{\mu}_{\sigma}{ + \frac{1}{2} D^2 (D_\nu h) }
+2D_{\mu}\left(R^{\mu \sigma}_{.\,\,.\,\,\nu \alpha}h^{\alpha}_{\sigma}\right)%-\nonumber \\
 +h^{\mu}_{\alpha}D_{\mu}R^{\alpha}_{\nu}
\nonumber }\\
 { + \frac{1}{2} R^\alpha_\nu\, D_\alpha h  -
D_{\mu}\left(h^{\alpha}_{\nu}R^{\mu}_{\alpha}\right) -
D_{\nu}\left(R^\alpha_{\beta} h_{\alpha}^{\beta}\right)   - \frac{1}{2} D_\nu D^2 h =
-2D_\mu T^{(1)\mu}_\nu .
}
\label{test1}
\ee

{
The first and the eighth terms in the l.h.s. of this equation are canceled, the second and the fifth terms are reduced to 
$D_{\mu}\left(R^{\mu \sigma}_{.\,\,.\,\,\nu \alpha}h^{\alpha}_{\sigma}\right)$ and we get:
%and the last term is equal to $R^\mu_\nu  D_\mu h$. }
\be
{
D_{\mu}\left(R^{\mu \sigma}_{.\,\,.\,\,\nu \alpha}h^{\alpha}_{\sigma}\right)
- R^{\sigma.\,.\,\mu }_{\,\,\nu\alpha}D_{\mu}h^{\alpha}_{\sigma}{+ \frac{1}{2} D^2 (D_\nu h) }
%-\nonumber \\
 +h^{\mu}_{\alpha}D_{\mu}R^{\alpha}_{\nu}
\nonumber }\\
 { + \frac{1}{2} R^\alpha_\nu\, D_\alpha h  -
D_{\nu}\left(R^\alpha_{\beta} h_{\alpha}^{\beta}\right)   - \frac{1}{2} D_\nu D^2 h =
-2D_\mu T^{(1)\mu}_\nu .
\label{test2}
}
\ee
Using commutation relations of covariant derivatives we find:
\be 
D^2 D_\nu h - D_\nu D^2 h = R^\alpha_\nu D_\alpha h.
\label{comm-h}
\ee

Eventually we arrive to
\be
{
h^{\alpha}_{\sigma} D_{\mu}R^{\mu \sigma}_{.\,\,.\,\,\nu \alpha}
+h^{\mu}_{\alpha}D_{\mu}R^{\alpha}_{\nu}-D_{\nu}\left(R^\alpha_{\beta} h_{\alpha}^{\beta}\right) + R^\mu_\nu  D_\mu h
=  -2D_\mu T^{(1)\mu}_{\nu}.}
\label{test3}
\ee
The first term here can be rewritten via the Bianchi identity as:
\be
h^{\alpha}_{\sigma} D_{\mu}R^{\mu \sigma}_{.\,\,.\,\,\nu \alpha} = 
h^{\alpha}_{\sigma}\left(D_{\nu}R^{\sigma}_{\alpha} -D_{\alpha}R^\sigma_\nu\right)
\label{bianchi}
\ee
leading to:
\be
h^{\alpha}_{\sigma}\left(D_{\nu}R^{\sigma}_{\alpha} -D_{\alpha}R^\sigma_\nu\right)+
h^{\mu}_{\alpha}D_{\mu}R^{\alpha}_{\nu} -
h_{\alpha}^{\beta} D_{\nu}R^\alpha_{\beta}   -  R^\alpha_{\beta} D_{\nu} h_{\alpha}^{\beta} +
R^\mu_\nu  D_\mu h  =
 -2D_\mu T^{(1)\mu}_{\nu}. 
%- \left( g^{\mu\alpha} T^{(1)}_{\alpha\nu}-h^{\mu\alpha} T_{\alpha\nu} \right).
%-2D_{\mu}\hat{T}^{(1)\mu}_{\,\,\,\,\,\,\,\,\nu}.
\ee
All the terms containing the derivatives of the Ricci tensor  in the equation above neatly cancel out and
finally we come to the following equation to be verified:
%After simplification we obtain the short equation
\be
R^\alpha_{\beta} D_{\nu} h_{\alpha}^{\beta}   -  R^\mu_\nu  D_\mu h =
2D_\mu  T^{(1)\mu }_\nu.
\label{transvers_eq}
\ee
}
Now we must check if  the l.h.s. and r.h.s. of the Eq. (\ref{transvers_eq}) are equal. To this end we 
will apply the following conservation conditions:
\be
\overline{D}_{\mu} \overline{T}^{\mu}_{\nu} = 0 \,\,\, {\rm and} \,\,\,
%\nonumber \\
D_{\mu}T^{\mu}_{\nu} = 0.
\label{condEMT}
\ee
So in the first perturbation order we obtain
\be
\overline{D}_{\mu} \overline{T}^{\mu}_{\nu} = D_{\mu}T^{\mu}_{\nu} + D_{\mu}T^{(1)\mu}_{\nu}+\Gamma^{(1)\mu}_{\,\,\,\,\,\alpha\mu}T^{\alpha}_{\nu}-\Gamma^{(1)\alpha}_{\,\,\,\,\,\nu\mu}T^{\mu}_{\alpha}.
\label{EMT0}
\ee
The third term in the r.h.s. of this equation is zero because the first-order perturbation corrections to the
Christoffel symbols has the following form 
\be
\Gamma^{(1)\mu}_{\,\,\,\,\,\nu\alpha} = \frac{1}{2}\left(D_{\nu}h^{\mu}_{\alpha}+D_{\alpha}h^{\mu}_{\nu}-
D^{\mu}h_{\nu\alpha}\right)
\label{Gamma-1-1}
\ee
and hence $\Gamma^{(1)\mu}_{\,\,\,\,\,\alpha\mu} =  D_\alpha h /2 $.

From Eqs. (\ref{EMT0}) and (\ref{Gamma-1-1}) we obtain
\be
&&D_{\mu} {T}^{(1)\mu}_{\nu} = \Gamma^{(1)\alpha}_{\,\,\,\,\,\nu\mu}T^{\mu}_{\alpha} - \Gamma^{(1)\mu}_{\,\,\,\,\,\alpha\mu}T^{\alpha}_{\nu} =
%D_{\mu}\tcmag{T^{(1)\mu}_{\nu}} =
\nonumber \\
%D_{\mu}\hat{T}^{(1)\mu}_{\nu}=
&&\frac{1}{2}\left(D_{\nu}h^{\alpha}_{\mu}+D_{\mu}h^{\alpha}_{\nu}-D^{\alpha}h_{\nu\mu}\right)\left(R^{\mu}_{\alpha}-\frac{1}{2}\delta^{\mu}_{\alpha}R\right)-\frac{1}{2}\left(R^{\alpha}_{\nu}-\frac{1}{2}\delta^{\alpha}_{\nu}R\right)D_{\alpha}h.
\label{EMT1}
\ee
Here we have taken into account the Einstein equations for the background metric (\ref{R-mu-nu-bcgr}).

After simple algebra we get: 
\be
D_{\mu}{T}^{(1)\mu}_{\nu}=\left( R^{\mu}_{\alpha}D_{\nu}h^{\alpha}_{\mu} - R^\mu_\nu D_\mu h \right) /2 .
\label{EMT}
\ee
Thus we see that both sides in equation (\ref{transvers_eq})
are equal. 
It means that  { condition~\eqref{Dpsi},  $D_{\mu}\psi^{\mu}_{\nu}=0$,} is compatible
with Eq.~(\ref{D2-h-mu-nu-h}).

{ Let us  calculate the trace of Eq.~(\ref{D2-h-mu-nu-h}).  We will find:
\be
D^2 h + 2 R^\mu_\nu h^\nu_\mu = 2 T^{(1) \mu}_{\mu}.
\label{trace-calc}
\ee
Seemingly, this equation is inconsistent with Eq.~\eqref{tr-main-eq}. However, if we use the relation between $T^{(1) \mu}_{\mu}$ and
 $g^{\mu \nu} T^{(1)}_ {\mu \nu}$, which we can obtain from Eq.~\eqref{bar-T-mu-nu}, we come to:
 \be
 T^{(1) \mu}_{\mu} = g^{\mu \nu} T^{(1)}_ {\mu \nu} - h_\mu^\nu \left(R^\mu_\nu - \frac{1}{2} \delta^\mu_\nu R \right)
 \label{T1-mu-mu} 
 \ee 
  and the consistency is restored. 

In the FLRW background $ R^{\mu}_{\alpha}D_{\nu}h^{\alpha}_{\mu} = R^\mu_\nu h^\nu_\mu = 0$ and the 
usual condition $T^{(1) \mu}_\nu = 0$ allows for $h = 0$. However, in arbitrary background it seems generally impossible  to impose both
conditions of covariant conservation of the source and its zero trace.  }

\section{Realistic metrics different from FLRW one \label{s-non-FLRW}}

Interesting deviations of the cosmological metric from the Friedmann one is induced by density perturbations
over the classical FLRW background. They are considered in detail in books~\cite{GR-2,Mukha,SW,CB-AD}, see also
Ref.~\cite{EA-AD-LR-Jeans}.
Assume that there exists a cloud of matter { with} energy density and pressure which are different from the average
cosmological ones. Generally speaking the cloud may be anisotropic but the impact of the additional term 
$h^{\alpha}_{\beta} R_{\alpha}^{\beta}$ (\ref{eq-mix-first}) would manifest itself even for isotropic distribution 
of matter in the cloud.
So, for simplicity, we confine ourselves to this case.
 
%maximum radius $r = r_m$
%initially constant energy density concentrated inside the maximum radius $r=r_m$. 
We choose Schwarzschild-like isotropic coordinates in which the metric takes the form:
\be
ds^2=Adt^2 - B\,\delta_{ij}\,dx^i dx^j\,, 
\label{ds-2}
\ee
where the functions $A$ and $B$ depend upon $r$ and $t$. The corresponding Christoffel symbols are:
\be 
&&\Gamma ^t_{tt}=\frac{\dot A}{2A}\,,\ \ \ \Gamma ^t_{jt}=\frac{\partial_j A}{2A}\,, \ \ \ 
\Gamma ^j_{tt}=\frac{ \delta^{jk} \partial_k A}{2B}\,, \ \ \ 
\Gamma ^t_{jk}=\frac{ \delta_{jk} \dot B}{2A}\,, \nonumber \\
&&\Gamma ^k_{jt}=\frac{ \delta^k_j \dot B}{2B}\,, \ \ 
\Gamma ^k_{lj}=\frac{1}{2B}(\delta^k_l\partial_j B + \delta^k_j\partial_l B - \delta_{lj}\delta^{kn}\partial_n B)\,.  
\label{Gammas}
\ee
The corresponding Ricci tensor is given by:
\be
R_{tt} &=& \frac{\Delta A}{2B} - \frac{3 \ddot B}{2B} + \frac{3 \dot B^2}{4B^2} + \frac{3 \dot A \dot B}{4AB} +
\frac{\partial^jA \partial_j B}{4B^2} - \frac{\partial^jA \partial_j A}{4AB}\, , \\
\label{R-tt}
R_{tj} &=& - \frac{\partial_j \dot B}{B} + \frac{\dot B \partial_jB}{B^2} + \frac{\dot B \partial_jA}{2AB}\, , \\
\label{R-tj}
R_{ij} &=& \delta_{ij} \left(\frac{\ddot B}{2A} - \frac{\Delta B}{2B} + \frac{\dot B^2}{4AB} -\frac{\dot A \dot B}{4A^2}
- \frac{\partial^kA \partial _kB}{4AB} +   \frac{\partial^kB \partial _kB}{4B^2}   \right)   \nonumber \\
&-&\frac{\partial_i\partial_j A}{2A} - \frac{\partial_i\partial_j B}{2B} +
\frac{\partial_i A\partial_j A}{4A^2} + \frac{3\partial_i B\partial_j B}{4B^2} +
\frac{\partial_i A\partial_j B + \partial_j A\partial_i B}{4AB}\, .
\label{R-ij}  
\ee  
Here and in what follows the upper space indices are raised with the Kronecker delta, $\partial^j A = \delta^{jk} \partial_k A$. 

The space derivatives of an arbitrary function of $r$ are equal to:
\be
\partial_i f = \frac{x_i}{r} f',\,\,\,\, \partial_i  \partial_i  f = \left( \frac{\delta_{i j} }{ r } - \frac{x_i x_j}{r^3} \right) f' + \frac{x_i x_j}{r^2} f'' ,
\label{dk-f}
\ee
where prime means differentiation over $r$. Thus $R_{ij}$ contains some other terms, except those proportional to $\delta_{ij}$,
and consequently the product  $h^{\alpha}_{\beta} R_{\alpha}^{\beta}$ is generally nonvanishing.

Usually perturbations are small, so both $A$ and $B$ weakly deviate from
unity and linear in $A$ and $B$ terms
dominate in  $R_{ij}$.  For a particular shape of the perturbations $(A - 1 )\sim r^2$ and $(B - 1) \sim r^2$, 
 as is considered in Ref.~\cite{EA-AD-LR-Jeans}, 
 and so $R_{ij} \sim \delta_{ij}$ and $h^{\alpha}_{\beta} R_{\alpha}^{\beta} = 0$. However,
it is not the general case.  Moreover, perturbations are known to rise as the cosmological scale factor at 
matter dominated regime,
and so the Ricci  tensor may become close by magnitude to its background value or even exceed it.

There are some more physically interesting metrics for which { the product $h^{\alpha}_{\beta} R_{\alpha}^{\beta} $ is 
nonzero}. One simple example is presented by
a collapse of dust-like matter, described e.g. in  book \cite{LL-2}.  This is the so-called 
pressureless Tolman solution. More general Tolman solution with non-zero pressure also possesses 
the same property.

\section{Conclusion \label{s-concl}}

 We have found that in an arbitrary curved background it is generally impossible to satisfy both standard 
conditions $D_\mu h^\mu_\nu = 0$ and $ h^\mu_\mu = 0$ on tensor perturbations. 
{As we have shown the
condition $D_\mu \psi^\mu_nu = 0$ (\ref{Dpsi}) can be imposed in any space-time.}
Of course tensor modes may
propagate in any background  but they would not be pure tensor modes or pure transverse modes. {
According to the equations derived in this work tensor and scalars modes are mixed and propagate together.

In some sense these phenomena resemble the  propagation of electromagnetic waves in plasma, when longitudinal modes are excited. 
Moreover, in inhomogeneous 
and anisotropic plasma propagating modes may be even more cumbersome. However, in high frequency
limit (eikonal approximation) we return to "normality". }

The discovered in this paper additional term in the equation for gravitational waves (GW) propagation 
in a background,  which differs from the FLRW one, may have an essential impact on the low frequency 
tail  of GW spectrum 
in particular,  of GWs which had been produced
during inflation, see Ref.\cite{StarGW,VerRubSazh}.  Their intensity at low frequencies might be
noticeably suppressed. 
{ Because of that, the strict limit on very long gravitational waves obtained from the CMB polarization 
data~\cite{CMB-GW} does not
necessarily implies that the traditional  inflation induced by a scalar field, inflaton, ~\cite{ADL-infl} is excluded.}
A study of this problem is in progress.

\section*{Acknowledgement}
We thank V.A. Rubakov for a very important comment and discussions.\\
The work was supported by RSF Grant 20-42-09010.

%\newpage

\end{document}